\documentclass[11pt]{article}
\usepackage{marginnote}

\usepackage{color}
\usepackage{amsmath}
\usepackage{amsthm}
\usepackage{comment}
\usepackage{marginnote}
\allowdisplaybreaks

\usepackage{array}
\usepackage{tabularx}
\usepackage{appendix}
\usepackage{makecell}
\usepackage{graphicx}
\usepackage{amssymb}
\usepackage[T1]{fontenc}
\usepackage[utf8]{inputenc}
\usepackage[raggedright]{titlesec}
\usepackage{blindtext}
\usepackage{commath}
\usepackage{cite}
\usepackage{caption}
\usepackage[textwidth = 7in]{geometry}
\usepackage{mathtools}
\usepackage[dvipsnames]{xcolor}
\usepackage{enumitem}
\usepackage{amsmath}
\usepackage{geometry}
\usepackage{lipsum}  
\usepackage{thmtools}
\usepackage{setspace}
\titleformat{\paragraph}[hang]{\normalfont\normalsize\bfseries}{\theparagraph}{1em}{}
\titlespacing*{\paragraph}{0pt}{3.25ex plus 1ex minus .2ex}{0.5em}
\makeatletter
\DeclareRobustCommand{\change}{%
	\@bsphack
	\leavevmode
	\color{magenta}%
	\@esphack
}
\DeclareRobustCommand{\stopchange}{%
	\@bsphack
	\normalcolor
	\@esphack
}
\makeatother

\stepcounter{secnumdepth}
\stepcounter{tocdepth}

\usepackage{hyperref}
\hypersetup{
  colorlinks   = true, 	
  urlcolor     = blue, 	
  linkcolor    = blue, 	
  citecolor   = magenta 	
}

\title{
	\vskip-1.3cm
Algebraic Solutions for Propagation of Elastic SH-Waves through an Isotropic Inhomogeneous Geometry\\
}

\author{ S. Zarrinkamar$^{1,2}$\footnote {saber.zarrinkamar@uva.es, saber.zarrinkamar@iau.ac.ir , ORCID: \href{http://orcid.org/0000-0001-9128-4624}{0000-0001-9128-4624}} 
	\\  [1ex]
	\small
	$^1$\,Departamento de F\'{\i}sica Te\'{o}rica, At\'{o}mica y \'{O}ptica, and Laboratory for Disruptive \\ 
		\small
		Interdisciplinary Science (LaDIS), Universidad de Valladolid, 47011 Valladolid, Spain\\
	         \small
	$^2$\,Departament of Basic Sciences, Garmsar Branch,
	Islamic Azad University, Garmsar, Iran
}

\begin{document}
	
	\maketitle

\begin{abstract}
The algebraic structure of the equation governing the propagation of an elastic SH-wave through an isotropic inhomogeneous layer surrounded by two homogeneous half-spaces is revisited. It is shown the problem under a quartic variation has a hidden $sl(2)$ symmetry based on which quasi-exact solutions are available via a completely analytical approach. This is of particular interest as the approach proposed for the corresponding tri-confluent Heun equation can be generalized to some other classes of geometries as well and the complicated considerations in case of Heun equations are absent.

\end{abstract}

\section{Introduction}
It looks quite logical that problems of mathematical engineering have some common points with those of mathematical physics simply due to the fact that such applied cases were initially considered as physical problems. However, and perhaps surprisingly, this relation has not been sufficiently addressed in some cases. In particular, in recent years there has been increasing interest in the study of Heun differential equation and ints confluent forms in both physics and engineering \cite {El-Jaick, Ronveaux, Ishkhanyan 2018, Ishkhanyan EPL 2016, Ishkhanyan scr}. 
The corresponding solutions, called Heun functions, have their own challenges as they are not as known as other simpler special functions of mathematical physics including Laguerre, Hermite, hypergeometric, etc.
Here, the propagation of an elastic SH-wave through an isotropic inhomogeneous layer is considered. The latter is an interesting problem in many areas of engineering and applied sciences and it has been considered with a variety of geometries and configurations. Some interesting examples with applications in composites are addressed in \cite {2013 Golub ref in 2019, book 2012, 2015 Vinh, 2019 main 1 Bednarika, 2021 j Low freq, 2021 Guo coating, 2023 Yang acta, 2024 brazilian, 2025 sound vib}. In our case, we consider a layer which is surrounded by two homogeneous layers and position-dependent shear modulus is assumed in the geometry. It is shown in \cite {2019 main 1 Bednarika} that the corresponding elastodynamic equation which is a partial differential equation can admit separation of variables and be transformed into an ordinary differential equation if a harmonic-type solution is assumed. \\       
To deal with the problem, we follow the idea of Lie-algebra \cite {Ushveridze, Turbiner88, Kamran 1990, Kamran 1991, Artemio94, Turbiner} as well as the Bethe-ansatz approach \cite {Book} as formulated by Zhang \cite {Zhang 2012, Zhang 2013, Zhang Annals 2016}, which works as well for Heun-like differential equations that are, roughly speaking, the most general form of a special function for which a closed form solution can be obtained under special conditions \cite {El-Jaick, Ronveaux, Ishkhanyan 2018, Ishkhanyan EPL 2016, Ishkhanyan scr}. The Bethe-ansatz technique can solve beyond the so called quasi-exact solvable problems formulated by Turbiner, Kamran, Artemio and Ushveridze \cite {Ushveridze, Turbiner88, Kamran 1990, Kamran 1991, Artemio94, Turbiner}, as well as Heun-like equations.\\
The structure of the present work is as follows. We first quote the most essential preliminaries of the elastodynamic equation for a composite system. The notations are preserved for a simpler comparison. Next, it is seen that the problem has a hidden $sl(2)$ symmetry which is confirmed by both the Bethe-Ansatz and Heun approaches. In the following section, derivation of eigenfunctions are commented on via the bethe-ansatz approach. The letter ends with some concluding remarks on the merits and failures of the present approach.
\section{Elastodynamic Equation}
Here, we follow the formulation with the notations of \cite {2019 main 1 Bednarika}, i.e. a plane time-harmonic elastic SH-wave propagating through an isotropic  inhomogeneous layer surrounded by two homogeneous half-spaces is considered. The  elastodynamic equation in terms of the displacement appears as
	\begin{equation} \label {PDE}
          \mu(z)\left[\frac{\partial^2u(x,z,t)}{\partial^2x}+\frac{\partial^2u(x,z,t)}{\partial^2z}\right]+\frac{d\mu(z)}{dz}\frac{\partial u(x,z,t)}{\partial z}=
\frac{1}{c_T^2(z)}\frac{\partial^2u(x,z,t)}{\partial t^2},
	\end{equation}
where Here $\mu(z)$ is the position-dependent shear modulus and $c_T(z)=\sqrt{\frac{\mu(z)}{\rho(z)}}$, with $\rho(z)$ being the mass density function. The shear stresses are
	\begin{equation} 
          \tau_{y,x}=\mu(z)\frac{\partial u(x,z,t)}{\partial x}, \ \ \tau_{y,z}=\mu(z)\frac{\partial u(x,z,t)}{\partial z}.
	\end{equation}
Considering the harmonic solution $u(x,z,t)=\tilde{u}(x,z)exp(-j\omega t)$ and the displacement as 
	\begin{equation} 
          \tilde{u}(x,z)=\hat{u}(z)exp(jk_Axsin\theta_{in}),
	\end{equation}
where $k_A=\frac{\omega}{c_A}$ is the wave number of the wave incoming from the homogeneous region A, $c_A$ denotes the transverse wave speed in the same
region, and $\theta_{in}$ stands for the angle of incidence, enables us to separate the variables  
	\begin{equation} 
          \frac{d^2U(s)}{ds^2}+\frac{1}{\eta(s)}\frac{d\eta(s)}{ds}\frac{dU(s)}{ds}+\left[K_0^2\frac{\xi(s)}{\eta(s)}-\varkappa^2\right]U(s)=0.
	\end{equation}
Here, $s=\frac{z}{l}$, $U=\frac{\hat{u}}{l}$, $K_0=k_0l=\omega l\sqrt{\frac{\rho(0)}{\mu(0)}}$, $\varkappa=k_Alsin\theta_{in}$, $\eta(s)=\frac{\mu(s)}{\mu(0)}$, $\xi(s)=\frac{\rho(s)}{\rho(0)}$
where $l$ is a characteristic length, $\eta(s)$ and $\xi(s)$ are called material functions. 
Introducing $\psi(s)=\phi(s)U(s)$, with $\phi(s)=\sqrt{\eta(s)}$, we obtain
	\begin{equation} 
          \frac{d^2\phi(s)}{ds^2}+V(s)\phi(s)=0.
	\end{equation}
In \cite {2019 main 1 Bednarika}, the authors consider a quartic profile of the form $V(s)=a_0+a_1(s-s_0)+2a_2(s-s_0)^2+a_3(s-s_0)^3-a_4(s-s_0)^4$, and after an exponential transformation (to ensure the validity of solutions) of the form $\phi(s)=exp\left(\frac{1}{2}\int{F(s)ds}\right)\varphi(s)$, 
where $F(s)=-2\sqrt{a_4}\left[\left(s-s_0)-\frac{a_3}{4a_4}\right)\right]^2+\frac{16a_2a_4+3a_3^2}{8a_4^{3/2}}$, as well as introducing $Q=\left(\frac{2\sqrt{a_4}}{3}\right)$, 
$\sigma=Q\left[(s-s_0)-\frac{a_3}{4a_4}\right]$, one obtains the equation \cite {2019 main 1 Bednarika}
	\begin{equation} 
          \frac{d^2\varphi(\sigma)}{d\sigma^2}-\left(3\sigma^2+\gamma \right)\frac{d\varphi(\sigma)}{d\sigma}+\left[\alpha+\left(\beta-3\right)\sigma\right]\varphi(\sigma)=0,
	\end{equation}
where
\begin{equation}
	\begin{aligned}
		\alpha=\frac{64a_0a_4^3+64a_2^2 a_4^2+32a_2a_3^3a_4+16a_1a_3a_4^2+3a_3^4}{64Q^2a_4^3},\\
		\beta=\frac{24a_4(a_1a_4+a_2a_3)+3a_3^3}{16a_4^{5/2}}, \ \ \ \gamma=-\frac{16a_2a_4+3a_3^2}{8Qa_4^{3/2}}.
	\end{aligned}
\end{equation}
It should be mentioned that a similar structure with the Pridemore-Brown equation has already been analyzed in \cite {Pridmore}.
\section{The Algebraic Structure of the Equation}
In this section, we comment on the algebraic approach to solve the equation in a quite analytical manner and without soft-ware help (at least for the primary states). The appendix provides the basic concepts of the idea as compact as possible for the unfamiliar readers, which are probably the majority of the engineering community.
As indicated in \eqref {coeff}, a second-order differential equation has a hidden $sl(2)-$algebraization, if and only if \cite {Zhang 2013, Zhang Annals 2016}
\begin{equation} \label {condition}
B_3=-2(n-1)A_4, \ \ \ \ \ \ C_2=n(n-1)A_4, \ \ \ \ \ B_1=-n\left[(n-1)A_3+B_2\right].
\end{equation}
The latter immediately yields in our case
\begin{equation} \label {from symmetry}
\beta=\beta_n=3(n+1).
\end{equation}
The dependence of the parameter on $n$ is often interpreted in quantum physics as the bound-state energies in the original idea of quasi-exactly solvable differential equations where the Schr\"{o}dinger equation's energy appears as $E=E_n$ where $n$ is the $state$. Here, however, it indicates that $\beta$, for each level, admits only a special discrete value. The latter is well understood in the parallel Heun function approach. To have a better idea, one can see Eq. (36) of \cite {2019 main 1 Bednarika}, which is easily derived here with this conceptual approach. 
It is now interesting to comment on the eigenfunctions as well. To this aim, we use the Bethe-Ansatz method as formulated by Zhang \cite {Zhang 2012, Zhang 2013}. If we propose a solution of the form
\begin{equation} \label{ansatz}
\varphi(\sigma)=\prod_{i-1}^n(\sigma-\sigma_i), \ n=1,2,\cdots , 
	\end{equation}
for the second-order Hamiltonian \eqref {Hamiltonian}, it can be found that \cite {Zhang 2012, Zhang 2013}
	\begin{equation}\label {General Form}
		\begin{gathered}
			-c_0-\sum_{i=1}^n\frac{Res(-c_0)_{\sigma=\sigma_i}}{\sigma-\sigma_i}=\left[n(n-1)a_4+nb_3+c_2\right]\sigma^2\\
                           +\left[\left(2(n-1)a_4+b_3\right)\sum_{i=1}^n\sigma_i+n(n-1)a_3+nb_2+c_1\right]\sigma\\
                           +\left(2(n-1)a_4+b_3\right)\sum_{i=1}^n\sigma_i^2+2a_4 \sum_{i<j}^n\sigma_i\sigma_j\\
                           +\left(2(n-1)a_3+b_2\right)\sum_{i=1}^n\sigma_i+n(n-1)a_2+nb_1.
		\end{gathered}
	\end{equation}
As a result, using  \eqref {ansatz} and \eqref {General Form} we may write 
	\begin{equation} \label {subs}
		\begin{gathered}
-\alpha=\sum_{i=1}^n\frac{1}{\sigma-\sigma_i} \sum_{ j\neq i}^n \frac{2}{(\sigma_i-\sigma_j)}
-\left(3\sigma^2+\gamma\right)\sum_{i=1}^n\frac{1}{\sigma-\sigma_i}+\left[\alpha+(\beta-3)\sigma) \right],
		\end{gathered}
	\end{equation}
which determines the $Res_{\sigma=\sigma_i}$ as
	\begin{equation} \label {Residues}
Res_{\sigma=\sigma_i}=\sum_{ j\neq i}^n \frac{2}{(\sigma_i-\sigma_j)}-\left(3\sigma_i^2+\gamma\right)=0.
	\end{equation}
The recent equation determines the roots $\sigma_i$ needed to obtain the eigenfunctions. Due to the fact that we have a constant term on the left-hand side, and a meromorphic function on the right-hand side, we may conclude that the second term on the right-hand side gives $nb_2+c_1=0$ which gives $\beta=\beta_n=3(n+1)$, already obtained in \eqref {from symmetry} from the $sl(2)$-algebraic structure of the equation.  
The constant term yields $\alpha+3\sum_{i=1}^n\sigma_i=0$. This is interpreted as a restriction, i.e., the approach works only for special values of $\alpha=\alpha_n$. This is not surprising since as already obtained in \eqref {from symmetry}, the approach works for certain values of the parameters. 

\section{Conclusion}
Some problems in engineering mathematics closely resemble some other problems in mathematical physics which arise in quantum or classical regimes. In particular, the idea of quasi-exactly solvable models, or conditionally-exactly solvable models, for which analytical solutions are available only in certain cases, are rather known in mathematical physics and in particular as a tool to obtain solutions of the one-dimensional Schr\"{o}dinger equation, or Schr\"{o}dinger-like equations. The purpose of this work was to introduce and apply the ideas of the Lie-algebraic approach, and namely the $sl(2)$ algebra, as well as the Bethe-ansatz approach to an elastodynamic problem. \\
It should be clearly stated that the approach, despite being rather simple and touchable in comparison with Heun approach, does have some problems including the assumption of resctriction among the engaged parameters. It does not report the solution in a closed form either. Nevertheless, it has its own merits and gives a quite good insight into the structure of the problem and can be as well used in checking numerical calculations. In addition, and most important of all, the idea used here works for a large class of geometries which yield in general the Heun equation and its confluent forms or even beyond them. However, it should be clearly mentioned that these two approaches are not equivalent. \\
In summary: any configuration in elastodynamic equations which results in a Schr\"{o}dinger-like equation with an effective potential similar to the ten classes of quasi-exactly solvable-cases can be investigated via the Lie-algebraic approach (see for example tables in \cite {Turbiner88}). The complete form of solutions, however, needs more calculations and physical concepts. Therefore, the Bethe-Ansatz approach was introduced here as well. Recalling that the general Heun equation has the form
\begin{equation*}
\Psi''(u)+\left(\frac{\gamma}{u}+\frac{\delta}{u-1}+\frac{\epsilon}{u-a}\right)\Psi'(u)+\frac{\alpha \beta u-q}{u(u-1)(u-a)}\Psi(u)=0,
\end{equation*}
a comparison with \eqref {Hamiltonian} reveals the power of Lie-algebraic and Bethe-Ansatz approaches. As the final point, it is worth noting that the interested reader can find quite summarized and applied connections between the two versions of Heun equation, i.e. with and without first-derivative, and the associated transformations in \cite {El-Jaick, Ronveaux}.

\section*{Aknowledgment}
S. Z. was supported by the Q-CAYLE project,funded by the European Union-Next Generation UE/MICIU/Plan de Recuperacion, Transformacion y Resiliencia/Junta de Castilla y Leon (PRTRC17.11), and also by project PID2023-148409NB-I00, funded by MICIU/AEI/10.13039/501100011033. Financial support of the Department of Education of the Junta de Castilla y Leon and FEDER Funds is also gratefully acknowledged (Reference: CLU-2023-1-05).

\appendix
\numberwithin{equation}{section}
\section{$sl(2)$ Algebra} 
Let us first comment on the jargon $sl(2)$ group. The letter $sl$ comes from the jargons  $special linear$group. Although the idea of $sl(2)$ algebra works for one complex variable, the idea is normally considered on real variable (simply because the original application, i.e. Schr\"{o}dinger equation, is defined as a derivative w.r.t a real variable in one dimension), and hence the formulation is alternatively denoted by $sl(2,R)$ or $sl_2(R)$. 
The building blocks of the algebra are $2\times2$ real matrices whose determinant is one.\\
In this appendix, the basic idea of quasi-exactly solvability is included as compact as possible to preserve the continuity and also to provide an insight for those who are unfamiliar with the idea, which will probably be the majority of the engineering community. The jargon quasi-exact comes from the fact that only some (and not all) states can be derived in such cases. This class goes beyond what is known The most general form of the second-order quasi-exactly solvable differential operator which can be expressed as an $sl(2)$ algebra is \cite {Ushveridze, Turbiner88, Kamran 1990, Kamran 1991, Artemio94, Turbiner} has the form
\begin{equation}
H_{QES}=A_{++}{J}_n^+{J}_n^++A_{+0}{J}_n^+{J}_n^0+A_{+-}{J}_n^+{J}_n^-+  A_{0-}{J}_n^0{J}_n^-+A_{--}{J}_n^-{J}_n^-+A_+{J}_n^++A_0{J}_n^0+A_-{J}_n^-+A,
\end{equation}
in which
\begin{equation}\label{GenerJJJ}
	\begin{aligned}
		J_n^+ &=z^2\,\frac{d}{dz}-n\,z ,\\
		J_n^0 &= z\,\frac{d}{dz}-\frac n2,\\
		J_n^- &= \frac{d}{dz}.
	\end{aligned}
\end{equation}
The generators, which are not unique, satisfy
\begin{equation}
[J^+_n,J^-_n]=-2J^0_n,  \ \ \ \   [J^{\pm}_n,J^0_n]=\mp J^{\pm}_n.
\end{equation}
As a result, it can be shown that $H_{QES}$ preserves the finite-dimensional space of polynomials of the form 
\begin{equation}
	\Phi_n (z)=\sum_{m=0} ^n{c_m z^m}.
\end{equation}
Using \eqref{GenerJJJ}, the operator $H_{QES}$ can be written as
\begin{equation} \label {Hamiltonian} 
H_{QES}=F_4(z)\frac{d^2}{dz^2}+F_3(z)\frac{d}{dz}+F_2(z),
\end{equation}
with
\begin{equation}\label{Coefficients}
	\begin{aligned}
	&	 F_4(z)=A_{++}z^4+A_{+0}z^3+A_{+-}z^2+A_{0-}z+A_{--},\\
	&	F_3(z)=A_{++}(2-2n)z^3+\left (A_++A_{+0}\left (1-\frac{3n}{2}\right)\right)z^2+\left (A_0-nA_{+-}\right)z+\left (A_--\frac{n}{2}A_{0-} \right), \\
	&	 F_2(z)=A_{++}n(n-1)z^2+\left(\frac{n^2}{2}A_{+0}-nA_+\right)z+\left (A-\frac{n}{2}A_0 \right), 
	\end{aligned}
\end{equation}
where all parameters (except the variable z) are constant. For the sake of simplicity, we write \eqref {Coefficients} as
\begin{equation} \label {coeff}
F_4(z)=\sum_{k=0} ^4A_kz^k, \ \ \ \ \ F_3(z)=\sum_{k=0} ^3B_kz^k, \ \ \ \ \ F_2(z)=\sum_{k=0} ^2C_kz^k.
\end{equation}


	\end{document}